**Wrinkles and Magnetic Flux Trapping in Graphite Nanoflakes: A Possible Source and Manifestation of Room-Temperature Superconductivity**

*Muhammad Saad, Sergey I. Nikitin, Dmitry A. Tayurskii, Roman V. Yusupov*

Institute of Physics, Kazan Federal University, Kazan 420008, Russia
E-mail: msaad797@gmail.com, Roman.Yusupov@kpfu.ru

Funding: The work was supported by a subsidy allocated to Kazan Federal University as part of a state assignment for scientific activities, project no. FZSM-2026-0021

The paper reports on a possible manifestation of local high-temperature superconductivity in graphite nanoflakes obtained from the bulk pyrolytic graphite by an extended grinding and subsequent annealing at 670 K in the air. Analysis of transmission electron microscopy (TEM) images shows that such treatment leads to a formation of high-density wrinkle-type defects at the basal-plane surface. The sample reveals the magnetic flux trapping – one of the key signatures of superconductivity – that persists up to 390 K and above. At the same time, both the as-ground (unannealed) and annealed in the vacuum samples manifest negligible flux trapping, and no wrinkles in their TEM images were found. The direct correlation between the appearance of high-density wrinkle arrays and the detection of the trapped magnetic flux above the room temperature strongly suggests that wrinkled regions serve a source of local high-temperature superconductivity in graphite.

**Graphical abstract:**

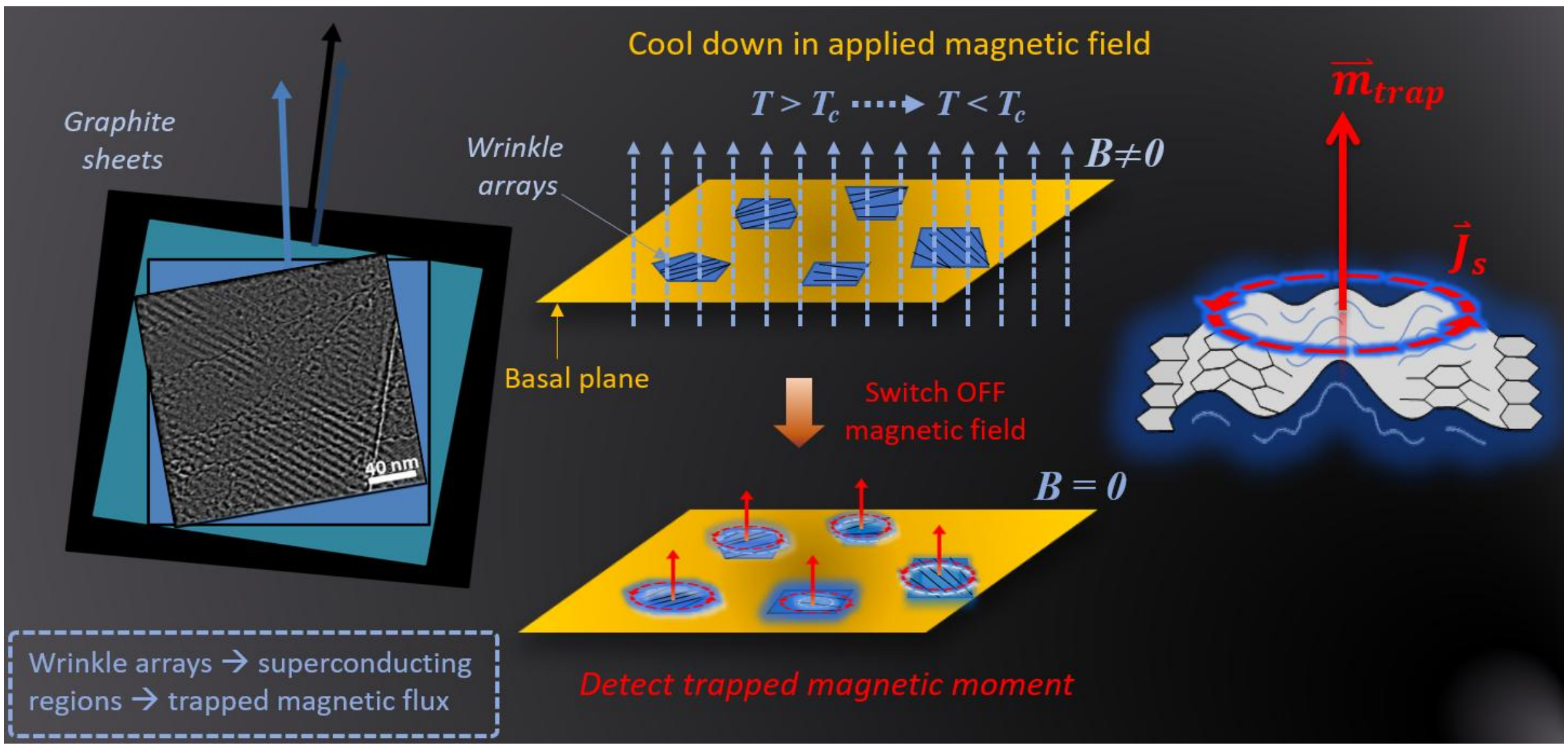

## 1. Introduction

The possibility of room temperature superconductivity in carbon-based materials has been a long-standing enigma in condensed matter physics, first proposed by W.A. Little in 1960s [1]. Early experimental observations of superconductivity were reported for intercalated graphites such as $C_8K$, $C_8Cs$ and $C_8Rb$ [2]. Later, in 1974, the first claims of granular room temperature superconductivity in graphite emerged, based on observations of Josephson tunneling-like current responses to applied magnetic fields. In that study, distorted carbon crystallites resembling graphite layers were sandwiched between two aluminum sheets followed by the annealing at 620°C [3]. Later on, several experimental hints were reported suggesting the existence of granular high-temperature superconductivity in different graphite-based samples. For instance, the sign of superconductivity (SC) in graphite has been explored under various conditions, including: quasi-2D interfaces [4–6], doping with water [7], phosphorus [8], sulfur [9], and in contact with alcohol [10] or in disordered structures [11-12]. In many cases, these superconducting states exhibited granular character when isolated superconducting regions contributed to the overall signal. Notably, a specific open magnetic hysteresis loop has also been observed in water-doped graphite powder in [7] that was considered as evidence for room-temperature superconductivity.

Recent studies suggest the possibility of granular, room temperature superconductivity (RTSC) in highly oriented pyrolytic graphite (HOPG) at ambient pressure, often linked to defect-rich interfaces or surface line defects formed during treatment. Thus, room temperature superconductivity was claimed [13] in the graphite wrinkled structure, where resistance drops were observed on multi-terminal transport measurements across an individual wrinkle. Resistance reduction with wrinkle development in graphene was also reported [14–17]. However, direct verification of the zero-resistance state in such small isolated areas remains challenging, and detecting the Meissner effect, another fundamental signature of SC, is not a straightforward task. In granular superconductors, transport measurements alone may not clearly approve the superconducting behavior, as the connections between the superconducting grains are often poor or absent. Instead, magnetization measurements provide a more reliable approach than the transport ones, as they are sensitive to persistent currents even in the absence of a fully percolated path. Therefore, trustworthy observation of the flux trapping in such granular structure could provide the researchers with valuable insights to the origin of high-temperature superconductivity in the graphite-based materials.

Recent reports on wrinkle studies in various 2D materials indicate that these objects are not merely the fancy morphological features but rather powerful tools for engineering future quantum electronic functionalities [18–19]. In particular, pioneering work from Rice University has demonstrated [17] that wrinkling in two-dimensional systems can precisely manipulate electron spin, enabling the formation of robust quantum states such as persistent electron spin, and potentially stabilizing localized superconducting phases [18, 19]. At the same time, theoretical predictions indicate that wrinkles form preferentially in restricted areas where layers are mutually twisted by small angles [20–22], suggesting that twisted layers and wrinkles act as structural motifs that promote collective electronic phenomenon [23]. Wrinkles at thin graphite or graphene basal planes commonly arise from strain relaxation [19], thermal fluctuations [24], or interlayer mismatch [25, 26], and they are often correlated with layer twisting [23-26]. Recent studies have shown that mild annealing in vacuum provides sufficient thermal energy for graphene layers to overcome metastable stacking configurations with mild ripples and enables stress-driven interlayer reconstruction to AB-stacking with flat surface and sharp billow-like defects [27]. It has also been demonstrated [28] that wrinkle formation in graphene represents an intrinsic strain-relaxation mechanism, where mechanically stored elastic energy is progressively released through the nucleation and evolution of out-of-plane deformation. Such structural defects including line defects and wrinkles can induce high local electron density of states (DoS) [29, 30]. Notably, while pristine graphite/graphene is not a superconductor, the introduction of dense arrays of line defects, particularly those generating significant strain can induce robust and sometimes room temperature local superconducting states [18-22, 28-32].

Direct observation of morphological defects on graphite surfaces also remains challenging due to its highly ordered stacked structure. Increasing layer numbers impose stronger out-of-plane constraints on carbon atoms, potentially suppressing wrinkle formation. Moreover, basal plane rearrangements are difficult to resolve microscopically, as they are often masked by overlapping spatial frequency features in electron microscopy data. Here, two-dimensional fast Fourier transform (2D-FFT) image filtering provides a unique approach to detect and isolate local structural rearrangements that are hardly visible using real-space imaging alone [33]. Such magic-strained regions, when subjected to external magnetic fields, may sustain persistent currents [19, 20]. From fundamental point of view, twisted structures and wrinkles represent a perfect model system for observation of various collective phenomena, such as superconductivity or Mott-insulator states [21]. Therefore, the aim of the present work is to prepare and study a graphite sample in a way promoting a formation of morphological features in which the superconductivity might be possible.

In this work, we adopt a simple and reproducible two-step process: an extended dry grinding of graphite followed by a controlled annealing. Remarkably, air annealing promotes the formation of high density of wrinkle areas at the graphite nanoflake basal plane, revealed by mesoscopic-scale analysis (TEM/FFT), while such corrugations remained essentially absent in vacuum-annealed and unannealed samples. Concomitantly, magnetization measurements reveal pronounced magnetic flux trapping exclusively in the sample bearing wrinkle networks, with trapped moments persisting up to at least $T \approx 390$ K. Wrinkle formation and flux trapping are directly correlated, revealing that basal plane structural features play a pivotal role in stabilizing superconducting-like behavior at and above the room temperature, offering a simple yet powerful route for engineering room-temperature quantum functionalities.

## 2. Results and discussion

### *2.1 Morphology evolution*

Scanning electron microscopy (SEM) was employed to check the morphological evolution of graphite samples – as-prepared, vacuum-annealed and air-annealed as shown in **Figure 1**. SEM images of the as-prepared sample (Fig. 1, a) exhibit a characteristic bulk-like graphite morphology: stacks of thin graphite sheets with their visible folding and mutual twisting, a direct consequence of an extensive prolonged grinding process. Sharp focusing is possible within only a part of the visible field; this indicates a significant out-of-plane sample spread.

The micrographs reveal different mesoscale responses of the graphite nanoflakes to the annealing environment, directly linking the annealing atmosphere to the type of generated structural features. A general feature of both annealed samples (**Figure 1, b and c**) is their dense out-of-plane sheet stacking (sample overall flatness). Observed folded edges look much steeper than those in the as-prepared sample. Step-like edges of the underlying graphite thin flakes are clearly manifested in the landscape of the top ones. Also, both annealed samples reveal notable basal-plane surface modification. It is very well pronounced for the air-annealed sample (**Figure 1, c**) though still can be resolved for the vacuum-annealed powder. Thus, both types of the annealing resulted in a relaxation of the graphite flakes so that they smoothly follow the underlying material landscape, and to a notable basal plane surface modification. Earlier, we have found that edge planes of the graphite flakes also experience modification, though air annealing resulted in improvement of the open flake edges while vacuum annealing decreased its crystal quality [34].

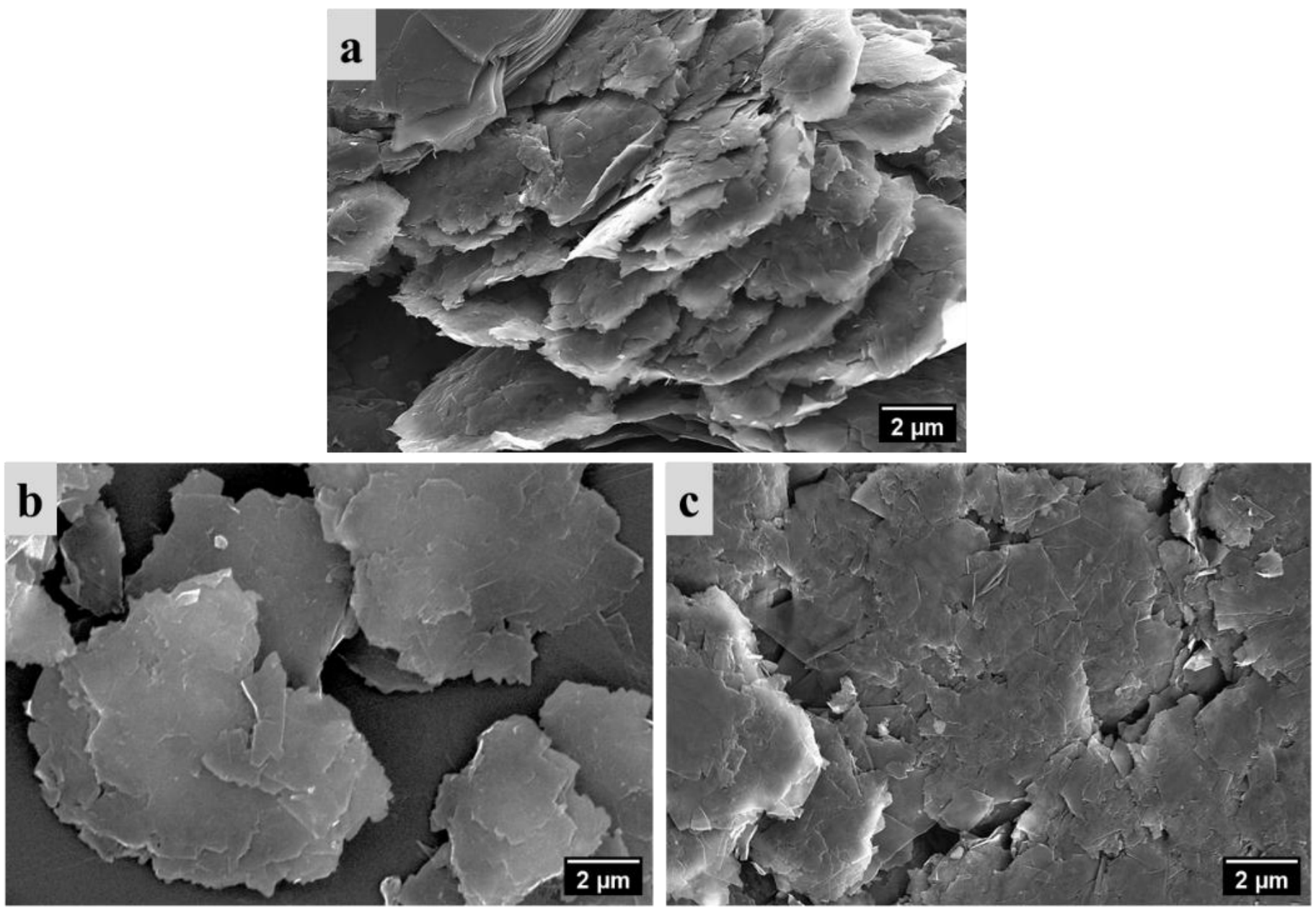

**Figure 1**: (a) Low-magnification SEM image of the graphite powders: (a) as-prepared, (b) vacuum-annealed, and (c) air-annealed samples.

The morphological evolution was further probed with low-magnification transmission electron microscopy (TEM) addressing the mesoscale sample (re)arrangements. Results for as-prepared, vacuum- and air-annealed samples are shown in **Figures 2 – 4**. Each figure is built over a low-magnification TEM-image of a given sample and has a universal composition. It includes the selected area electron diffraction (SAED) from a spot indicated by the dashed circle, 2D-FFT intensity distribution from a region shown with the red rectangle, and a set of higher-magnification reconstructed TEM-images with high spatial frequencies filtered out by means of 2D-FFT/filtering/2D-IFFT procedure [33]. Results for distinct areas are indicated by color-line boxes. Extra TEM images of other regions of the same studied samples as well as TEM image evolution with FFT-filtering are shown in Figs. S2-S4 of the Supplementary Material.

One can clearly see that as-prepared flakes are characterized by a complex pattern of Moire' fringes (**Figure 2**), which is a real-space manifestation of the interference between rotationally misaligned (twisted) graphite layers. SAED panel indeed reveals multiple diffraction patterns from twisted overlapping graphite sheets. Moreover, the 2D-FFT panel as well as the reconstructed TEM image of region 1 manifest several different low spatial frequencies. Reconstructed TEM images of regions 2 and 3 also reveal multi-domain Moire' patterns, indicating a complex landscape of variably twisted graphite layers distributed across the flake basal-plane. Thus, the applied reconstruction of TEM images from filtered 2D-FFT data is a simple and powerful approach to reduce the masking noise in the space domain to observe directly mesoscopic-scale surface rearrangements [33, 34].

Upon annealing the sample in vacuum (**Figure 3**), the 2D-FFT pattern deformed into diffuse halo, indicative of amorphous-like sample structure. The corresponding reconstructed TEM-image regions show overlapping small graphite pieces with sharp edges.

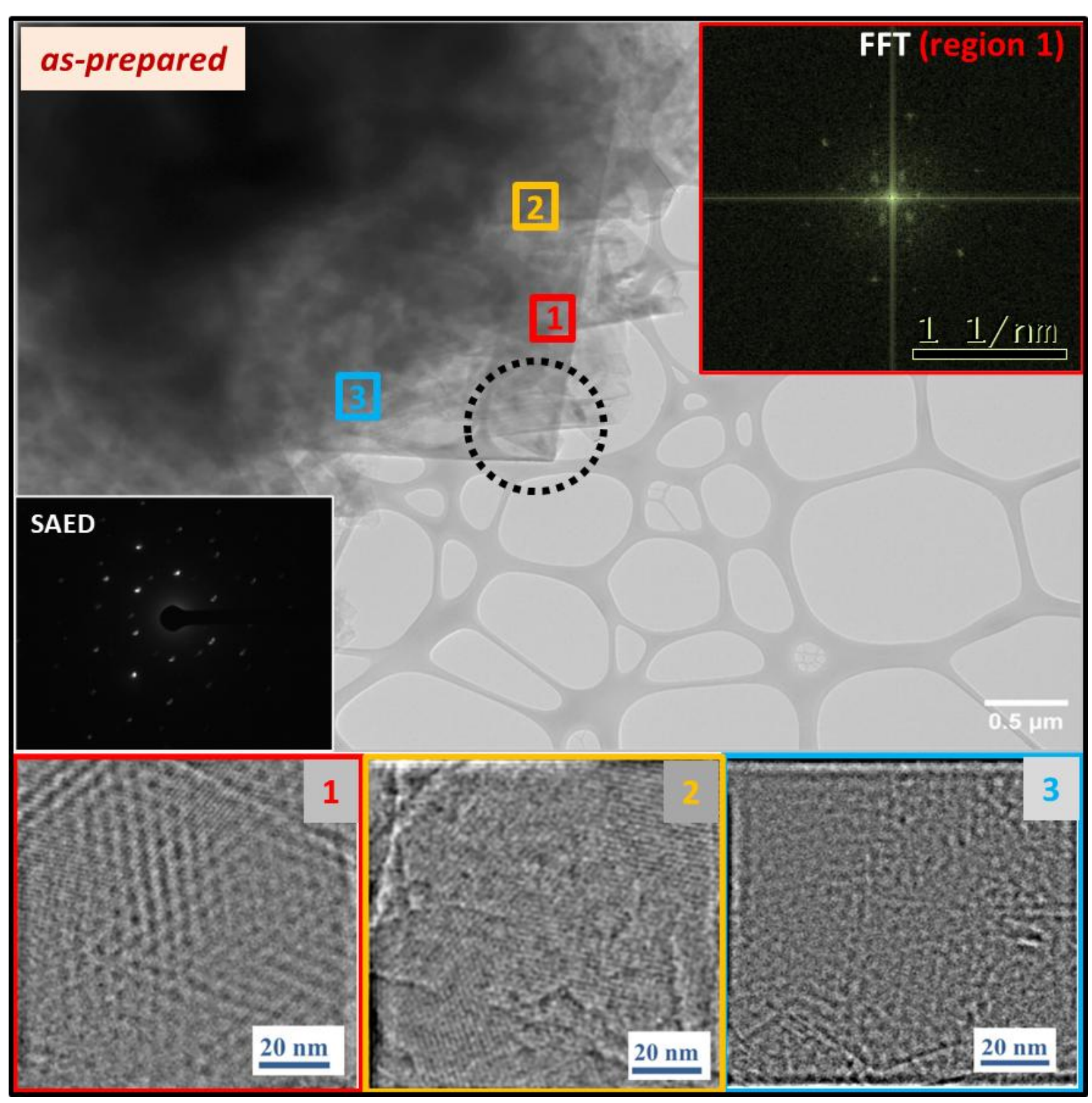

**Figure 2**: TEM analysis of the as-prepared graphite nanoflake sample. The 2D-FFT pattern (upper right inset) was obtained from the region marked by the red box in the TEM image, while the SAED pattern (bottom left inset) is collected from the region indicated by the dashed circle in the TEM image. Lower panels are reconstructed TEM micrographs of regions 1 – 3 (see text).

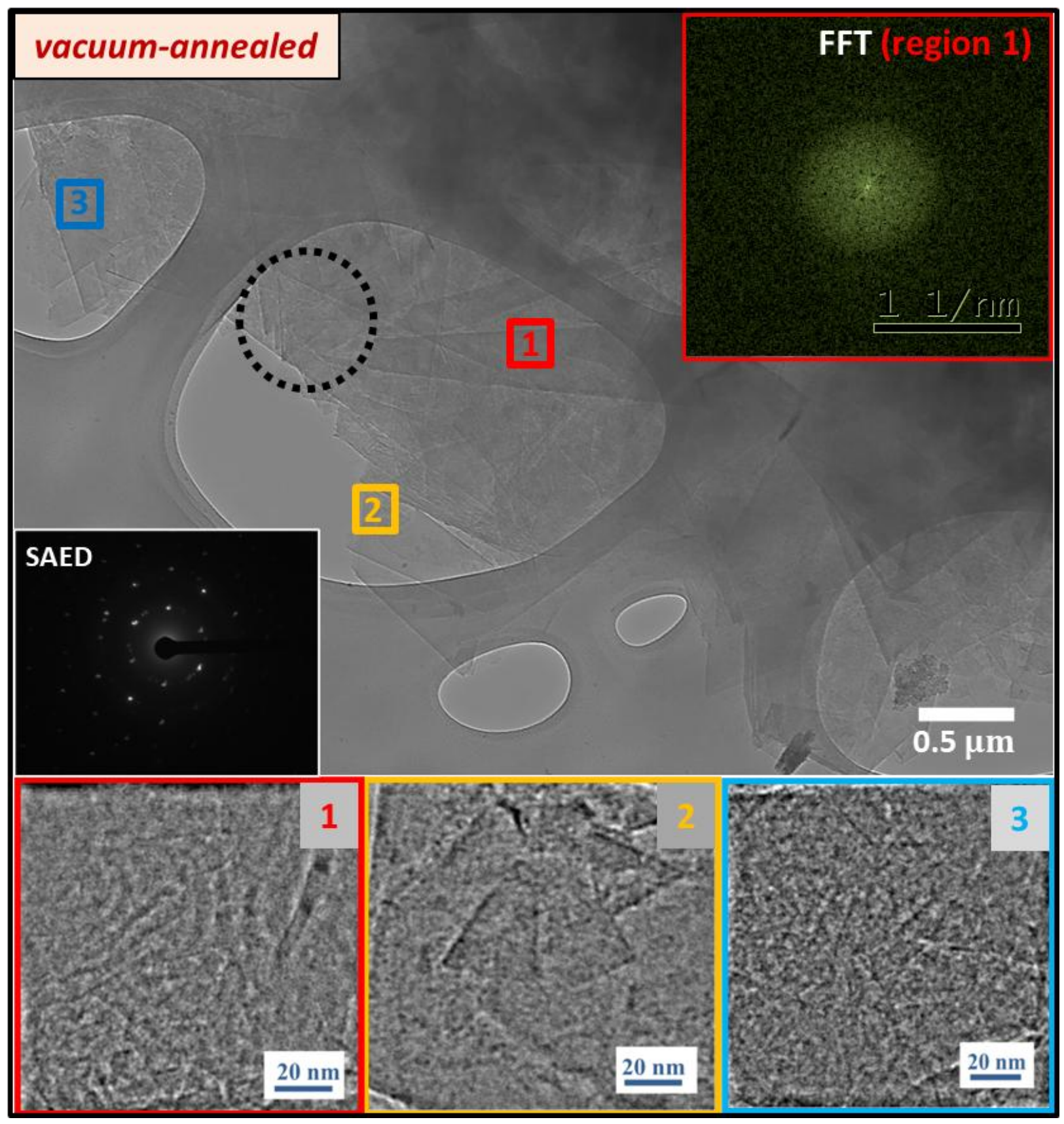

**Figure 3**: TEM analysis of the vacuum-annealed graphite nanoflake sample (same figure composition as that of **Figure 2**).

In contrast, upon mild annealing in the air (**Figure 4**), this metastable landscape of twisted interfaces undergoes a fundamental reorganization. The Moire' patterns vanish, giving way to an extended line networks of nearly parallel wrinkles and occasional defects healing. The 2D-FFT pattern containing additional intensity maxima (arrow indicated) proximal to the central spot, a direct consequence of low-frequency spatial modulations characteristic of new out-of-plane objects (wrinkles). By filtering the noise in the frequency domain as used in [33], the reconstructed TEM image discloses the emergence of dense nearly parallel quasi-periodic features distributed across the basal planes reflecting strain-relief-driven corrugation. Such defects, which extend across the areas of tens to hundreds of nanometers in size and possess the period of few to tens nanometers (see Figure 4), can be identified as wrinkles formed in the course of temperature-driven strain relief. This morphological transformation indicates that the stored strain energy is released through a collective buckling into a coordinated, out-of-plane surface corrugation.

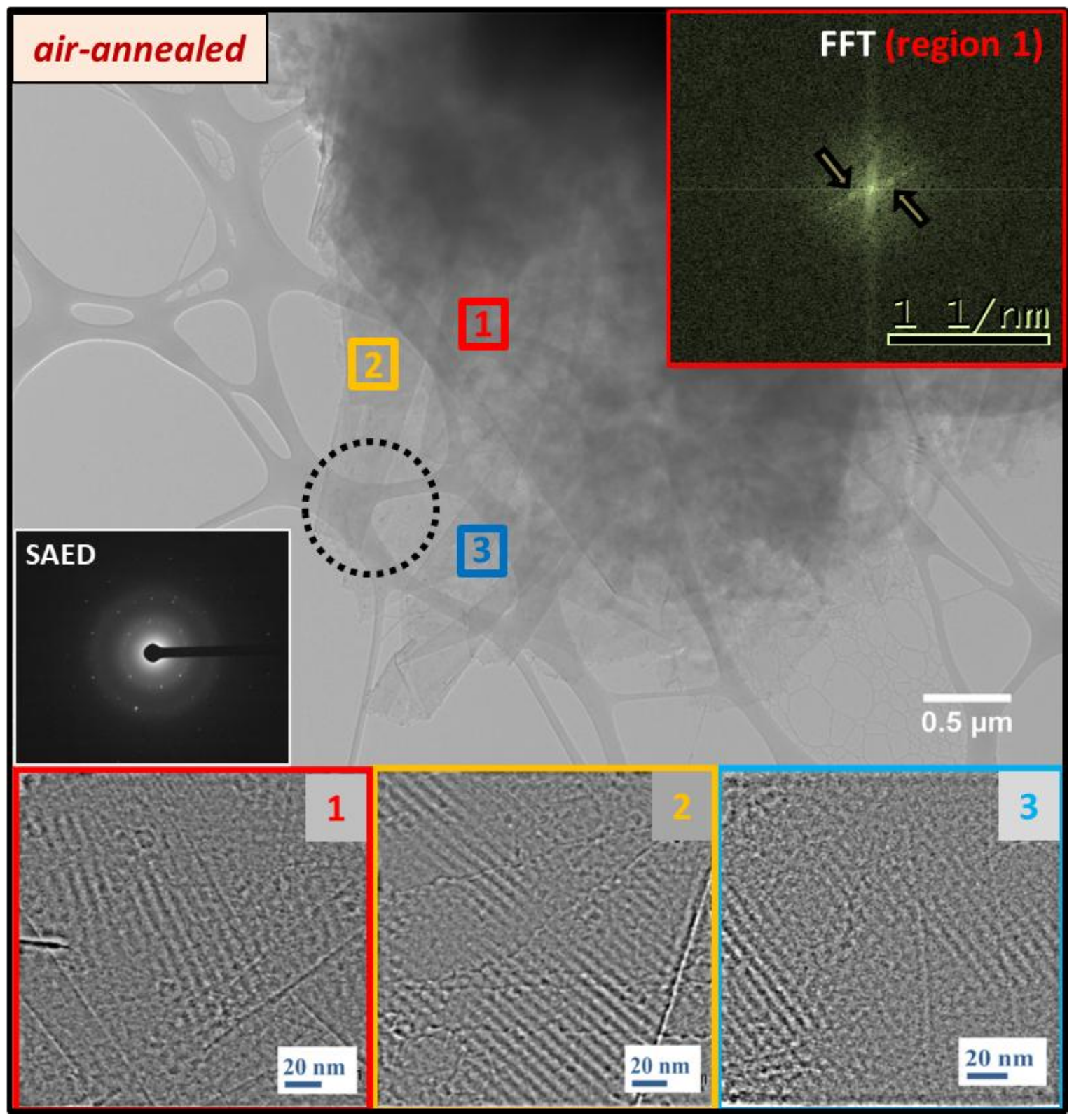

**Figure 4**: TEM analysis of the air-annealed graphite nanoflake sample (same figure composition as that of **Figure 2**).

In previous studies of graphite/graphene samples, attention has largely been focused on Moiré patterns arising from rotational misorientation between adjacent layers that are clearly revealed in TEM images like in Figure 2 [4-6, 22]. However, closer inspection of the SEM micrographs (see, e.g., Figure 1, a) also reveals the presence of large-scale undulations or swellings, which were not discussed earlier. Such features have typical lateral dimensions on the order of several tens to few hundred nanometers.

Here, we tentatively present a simple mechanism of wrinkled regions formation in the course of the applied sample preparation procedure (**Figure 5**). Prolonged dry grinding introduces substantial mechanical strain, rotational disorder, and metastable stacking configurations into the graphite nanoflakes, thereby mechanically activating the material. The bubble-like defects schematically shown in the upper part of **Figure 5** are formed due to a shear-type impact on individual flakes imposed to it between the mortar and pestle surfaces in the course of grinding. During subsequent annealing in the air, the stored elastic energy is progressively released as thermal activation enhances interlayer sliding, structural rearrangement and local stacking relaxation. Similar stress-driven reconstruction has been reported for bilayer graphene [27], promoting the evolution of strained ripple- or billow-like structures [26, 27]. In our case, the mechanically distorted multilayer graphite instead relaxes via a development of wrinkle arrays at the basal plane (Figure 5), providing an efficient pathway for accommodating residual strain while preserving the continuity of the upper graphene layers.

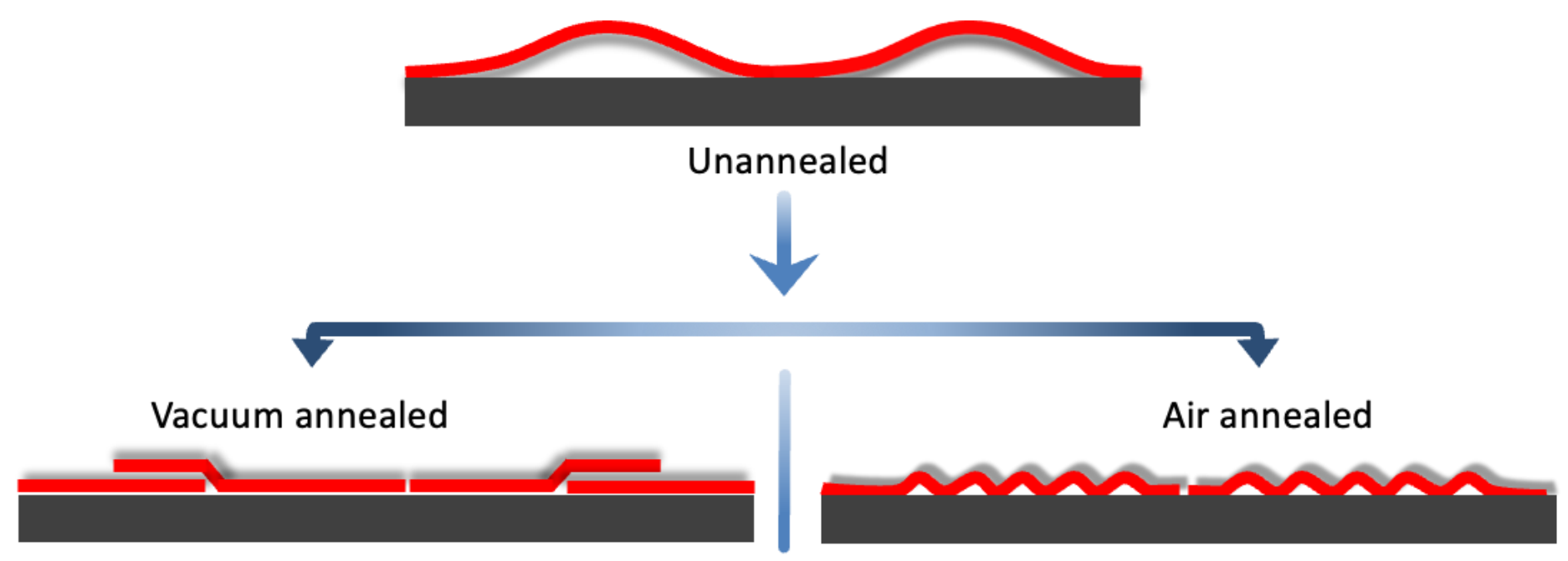

**Figure 5:** Schematic view of morphology transformation with the mild annealing at 670 K of the finely ground graphite flake surface (top) under vacuum (left) and air (right) atmospheres.

Another although very similar path towards wrinkle formation was proposed in Ref. 33 and represent essentially a strain-relief process within initially decoupled and twisted graphite sheets. The Moire' patterns visible in the as-prepared graphite flakes (**Figure 2**) confirm the presence of a long-range, coherent rotational misalignment between layers. This twist introduces significant in-plane shear strain. Similar to the above, thermal energy enables the system to relax this strain. The most energetically favorable pathway for this relaxation is through the out-of-plane buckling, leading to the formation of wrinkles [33]. These wrinkles are the remnants of the strain fields that were induced in graphite flakes during the extended grinding. The transition from a uniform Moire' pattern to a network of wrinkles demonstrates that the annealing process disrupts the long-range twist coherence, localizing and releasing the strain into specific, well-defined structural features. Thus, the observed wrinkles can be interpreted as the final, frozen signature of the pre-existing twisted regions.

Interestingly, strain release mechanisms are different under vacuum conditions and in the air at a temperature of 670 K. It looks that in the course of the annealing in vacuum, thin graphite/graphene sheets tear in strained regions and lay one edge onto another (**Figure 5**, left bottom sketch). A similar observation of overall sample flattening under annealing in vacuum was reported for graphene samples [27]. Under annealing in the air, topmost graphite layers do not tear, remain continuous, but alter their morphology, forming arrays of near-parallel mesoscopic-scale folds/wrinkles (**Figure 5**, bottom right). These observations highlight the critical role of the annealing environment in governing the strain relaxation pathway, given that extensive dry grinding introduces generative stress, defects and rotational disorder in graphite nanoflakes.

*2.2 Magnetization measurements*

Further, static magnetic properties of the prepared graphite nanoflake samples were studied with the vibrating sample magnetometry (**Figure 6**), by measuring $M(H)$ curves following a controlled cooling protocol. First, each sample was cooled from 300 K down to 10 K under zero magnetic field to obtain the zero-field-cooled (ZFC) response. The ZFC measurement corresponds to the virgin magnetization loop, recorded by sweeping the magnetic field in the sequence of 0T → +1T → –1T → +1T. The intrinsic diamagnetic susceptibility $\chi_{dia}$ was extracted from high-field linear slope of the ZFC $M(H)$ curves measured at $T = 10$ K. The as-prepared sample exhibits $\chi_{dia} = -9.87 \times 10^{-6}$ emu g$^{-1}$ Oe$^{-1}$, consistent with the orbital diamagnetism of the HOPG graphitic systems. Vacuum annealing results in a slightly reduced $\chi_{dia} = -9.68 \times 10^{-6}$ emu g$^{-1}$ Oe$^{-1}$, suggesting that thermally induced structural rearrangement moderately relaxes the electronic structure without significant altering the underlying diamagnetic character. In contrast, the air-annealed sample shows a substantially enhanced $\chi_{dia} = -1.07 \times 10^{-5}$ emu g$^{-1}$ Oe$^{-1}$.

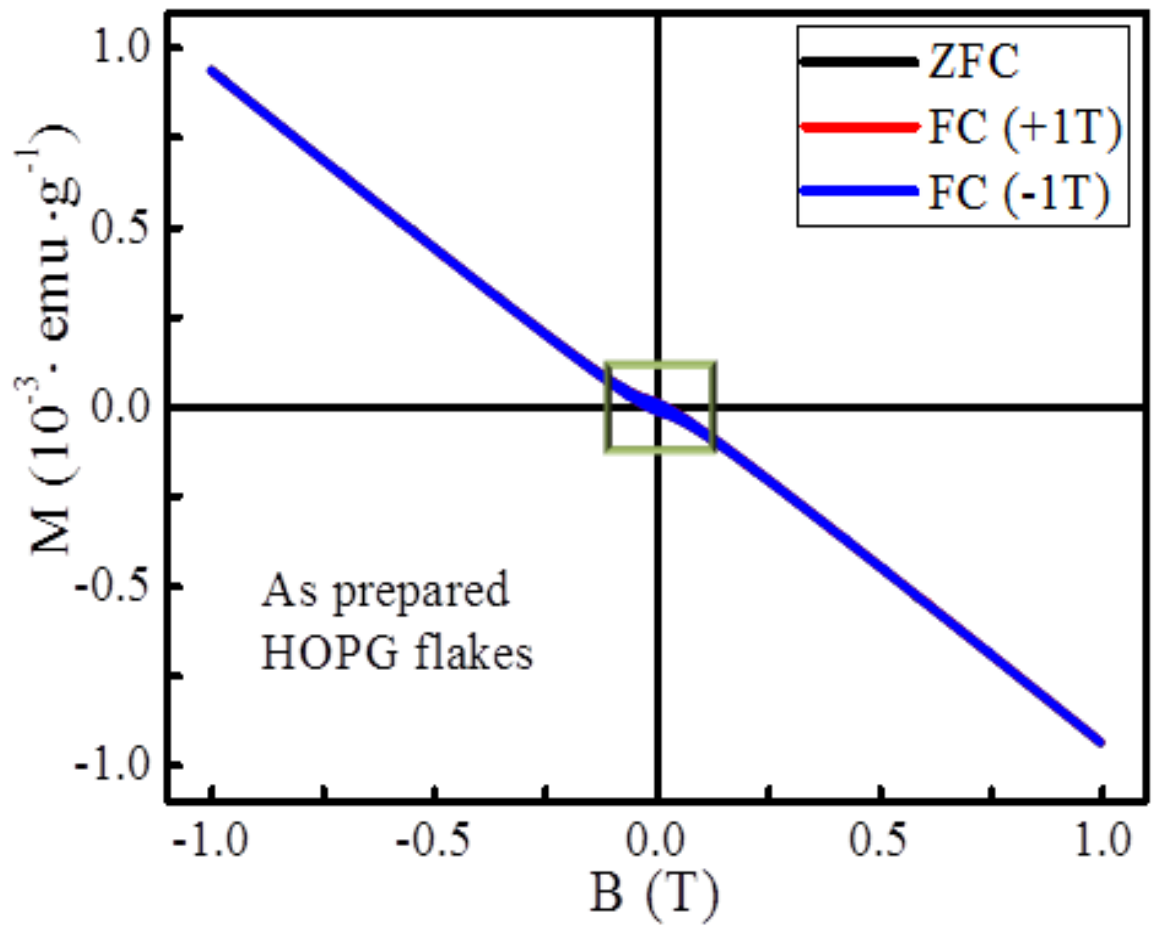

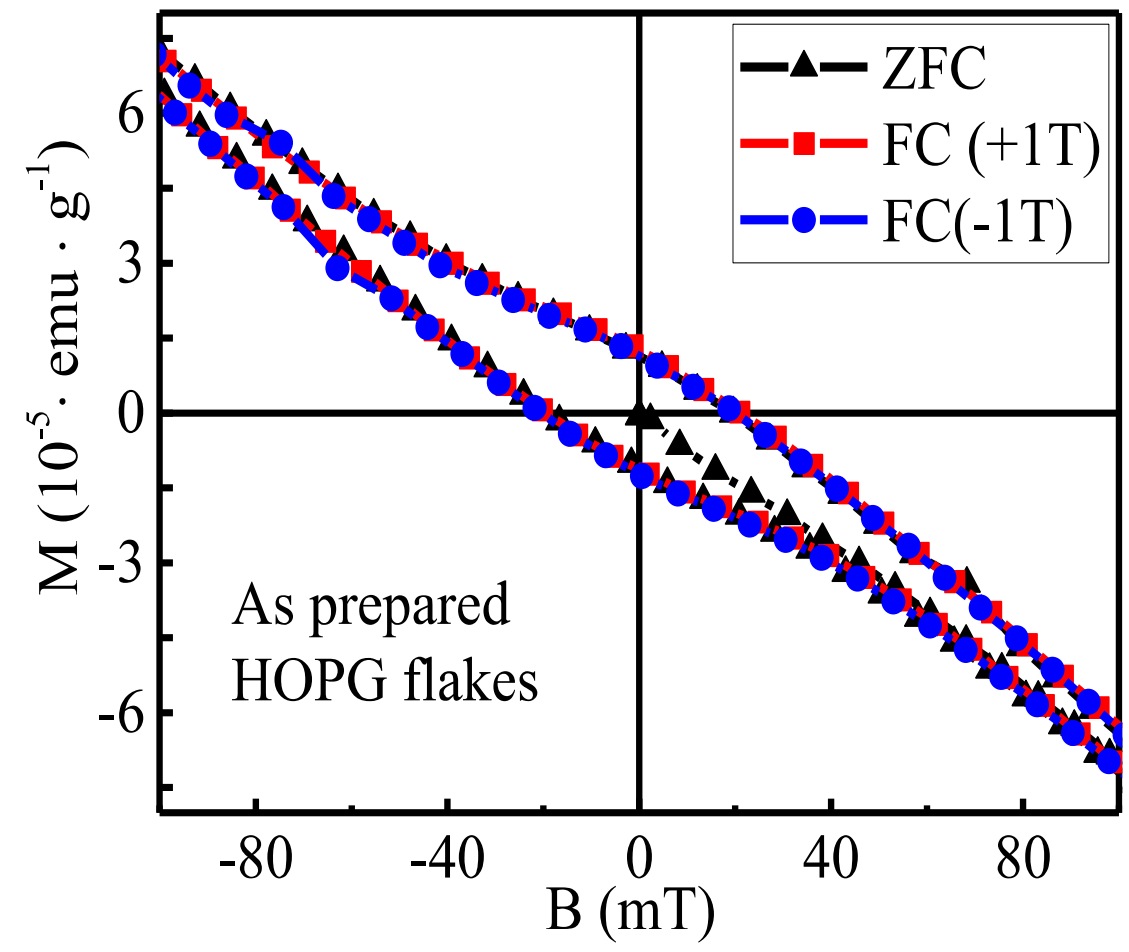

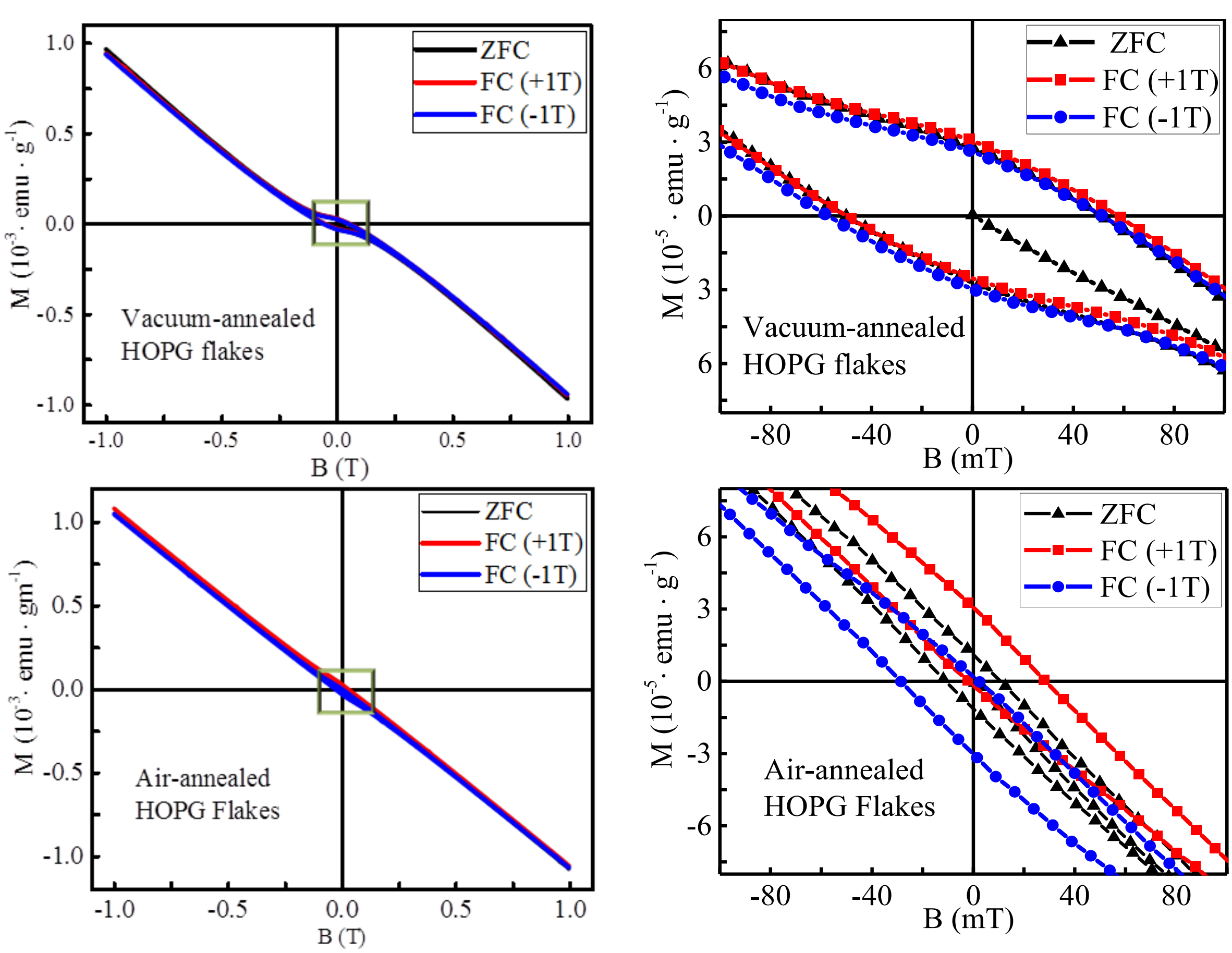

**Figure 6:** Magnetization $M(H)$ dependences at $T = 10$ K for the as-prepared, vacuum-annealed and air-annealed samples. For each panel, the ZFC $M(H)$ curve is shown together FC ones with a field of + 1 T (red) or – 1 T (blue) was applied during cooling. Panels on the right show magnified views of the low-field regions (± 100 mT) indicated by rectangles on the left.

These modifications are consistent with TEM data. Thus, reducing the lateral flake size on its tear under vacuum annealing restricts the continuous area for current induction and leads to a decrease of diamagnetic susceptibility. The air-annealed sample on the other hand exhibits flattened and continuous basal planes and experiences defects healing, which enhance electron delocalization and promote extended current pathways, leading to the observed increase in diamagnetic susceptibility.

In further measurement, two field-cooled (FC) curves were obtained by cooling the same sample from 300 K to 10 K under an applied field of either +1T or –1T, followed by recording the magnetization loops using the same sweep protocol. The resulting $M(H)$ curve families for each sample manifest pronounced differences among the samples. For the as-prepared nanoflakes, all the three curves (ZFC and the two FC curves) are symmetric with respect to the origin, and FC loops essentially overlap with the ZFC loop. In the vacuum-annealed sample, the curves remain almost symmetric, although with a slightly enhanced coercivity along with small vertical shifts. In contrast, the air-annealed sample exhibits a clear asymmetry between the ZFC and FC curves: the FC (+1T) and FC (–1T) loops are clearly displaced vertically in the direction of cooling fields, demonstrating a pronounced field-history-dependent offset. This behavior becomes especially evident in the zoomed region around ±100 mT, shown on the right for each sample (**Figure 5**). To address the possible ambiguity between a vertical displacement (trapped moments) or a horizontal shift (exchange bias), we performed a manual translation analysis of the FC hysteresis loops relative to the ZFC loop, as detailed in Figure S5 of the Supplementary Material. When FC curves were shifted vertically, they became fully overlapped and coincident with the ZFC curve demonstrating that the apparent displacement can be completely accounted for by a constant vertical offset. In contrast, applying horizontal translation to the FC curves failed to restore the symmetry, and the curves remained mismatched relative to the ZFC curve. This comparative analysis unambiguously confirmed that the observed displacement originates from trapped magnetic moments (i.e., a vertical shift), rather than from an exchange-bias-type horizontal field shift.

Trapped flux is one of the most distinct experimental manifestations of superconductivity. The last was reported in graphite several times [7, 10, 12, 38-39], and the recent work by Kopelevich and co-authors [13] demonstrated with the direct 4-probe measurements that a single wrinkle can become locally superconducting. Importantly, flux trapping is observed namely in our air-annealed graphite sample for which self-arrangement into wrinkled areas takes place as shown in **Figure 4**. Indeed, two other studied samples, the as-prepared and the vacuum-annealed ones, manifest neither wrinkled areas nor flux trapping.

According to the results of Kopelevich et al [13], wrinkle-associated local superconductivity persisted well above the room temperature. Therefore, we studied temperature dependence of the trapped magnetic flux. Here, **Figure 7(a)** shows the remnant magnetic moment after field cooling (FC) from 300 K to 10 K in an applied field of +1 T or −1 T and consequent field zeroing either immediately or after applying an opposite direction field. Irrespective of the origin of the observed magnetic hysteresis, trapped magnetic moment is defined as an ordinate of its center therefore was calculated as $M_{trap} = \left(M_r^+ + M_r^-\right)/2$, where $M_r^+$ and $M_r^-$ are remnant magnetizations after field removal from +1 T and –1 T, respectively. After field zeroing at 10 K, the remnant magnetization was recorded on warming. Thus, two quantities $M_r^+$ and $M_r^-$ were measured for each of the FC procedures, in +1 T and –1 T.

Trapped moments are extracted as an average of $M_r^+$ and $M_r^-$ and are shown in **Figure 7(b)**. The clear separation between the two branches demonstrates stable persistent currents that reveal themselves up to a temperature of at least 390 K, consistent with the vertical shift observed in the *M*(*H*) curves. The observed behavior reflects a robust flux-trapping mechanism in the air-annealed sample, likely originating from wrinkle-induced local current loops absent in the as-prepared and vacuum-annealed nanoflakes.

The trapped moment of $\approx 2.1\times10^{-5}$ emu/g is established upon cooling below 350 K. On warming, this moment relaxes gradually but persists up to $T > 390$ K. The cooling-field direction controls the sign of the trapped moment: +1 T cooling yields positive $M_{trap}$, and –1 T yields negative $M_{trap}$, confirming that the effect is a direct consequence of the field present during the superconductivity onset. This robust field-polarizable trapped flux that lives up to 390 K, is observed exclusively in the sample that possesses the wrinkled regions providing a direct structure-property link. Partial drop in trapped moment on warming indicates an occurrence of the critical temperature distribution within superconducting areas. Probably, different wrinkle arrangements possess different $T_c$'s. Areas with the $T_c > 390$ K are present in the sample. Our results are consistent with the assignment of the zero-resistance state to a wrinkle at the surface of graphite [13]. Magnetic flux trapping can be considered as another proof of above-the-room temperature superconductivity manifestation in specially-annealed graphite.

Importantly, a described set of observations have been reproduced in several independently prepared samples with qualitatively identical though quantitatively slightly different outcome [39, 40]. One of the earlier investigated samples was contaminated with cobalt oxide that significantly altered the integral magnetic properties of the samples. It complicated the interpretation but still allowed to reveal clearly the magnetic flux trapping in air-annealed specimen [39]. In the present study, XRD, XPS, EDS and AAS measurements (see Supplementary Material) reveal no detectable secondary magnetic phases, while the observed flux trapping appears exclusively in the air-annealed samples exhibiting wrinkle networks at the graphite basal plane. Thus, the absence of comparable behavior in both the as-prepared and vacuum-annealed samples, despite their similar chemical composition, suggests that the trapped magnetic moments are primarily associated with the wrinkle-induced structural state rather than with extrinsic magnetic impurities.

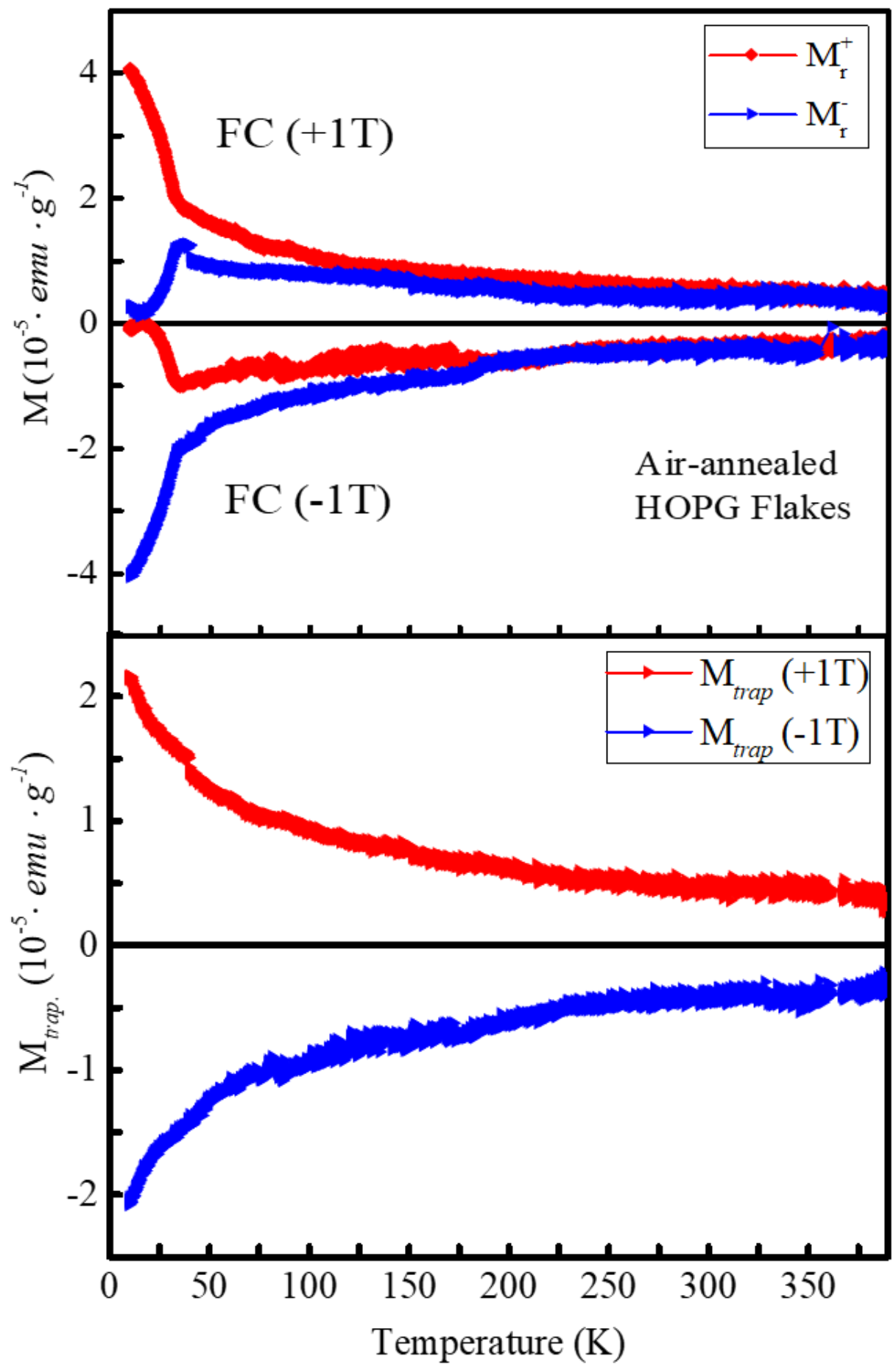

**Figure 7**: Temperature dependences of the remnant magnetization (a) and the trapped magnetic moment (b) of the air-annealed graphite nanoflake sample.

In the recent work on room temperature superconductivity in the HOPG sample decorated with wrinkles [13] a critical current $I_c$ was revealed, governed by the normal state resistance, using multiterminal transport measurements. The authors of [13] further reported an enhancement of diamagnetism and a suppression of the ferromagnetic response in room temperature $M(H)$ measurements following sample annealing. Their finding, supported by a theoretical model of Josephson-coupled superconducting granules embedded within wrinkles, demonstrate that such linear defect arrays can act as tailored hosts for macroscopic quantum coherence. A crossover field, $B_x \approx \pm$ 35 mT, has been identified in [13], where the behavior of the critical current $I_c(T, B)$ changes from $dI_c/dT < 0$ to $dI_c/dT > 0$, which is well known behavior for type-II superconductors and is mostly observed either in the vicinity of the upper critical field $B_{c2}$ *(T)* or just below the Abrikosov vortex lattice melting phase transition. In our case, direct resistance measurements in such granular samples are often ambiguous due to the percolative nature of the superconducting path, contact resistance and the potential masking of the zero resistance by serial conduction through normal graphite matrix. Whereas magnetic flux trapping provides an unambiguous and contact-free signature of the persistent currents. The wrinkles trap the Abrikosov vortices induced by the external magnetic field. When the field is removed, flux is trapped (remnant magnetization). This behavior is consistent with the critical state model for type-II superconductors [41], wherein magnetic flux penetrates the sample in the form of Abrikosov vortices (quantized flux tubes) that become pinned by structural defects. In this case, the wrinkle network acts as an effective pinning landscape. Within each strained wrinkle segment or at wrinkle intersections, the local electronic structure is modified. These wrinkled regions or superconducting islands are weakly coupled to each other through adjacent normal graphite matrix. Upon field cooling through the transition temperature ($T_c \cong 390$ K), magnetic flux is trapped as vortices become immobilized by the wrinkled topology. The resultant remnant magnetization ($M_{\text{trap}} \neq 0$ at $B = 0$) is the macroscopic manifestation of these pinned vortices. In contrast, the absence of flux trapping in vacuum-annealed samples (which exhibit no wrinkles) provides a crucial control, confirming that the wrinkled topology is essential for both vortex pinning and the superconductivity itself. In total, the observation of the trapped flux up to 390 K places the lower bound of the superconducting transition well above the room temperature. Our findings are also consistent with the rapidly emerging view that multilayer graphitic systems provide a fertile platform for superconductivity [42]. Thus, the observed wrinkle networks may provide an alternative structural route for local modification of the electronic structure and promoting superconducting regions in graphite.

**3. Conclusion**

The chase of room-temperature superconductivity represents a fundamental challenge in condensed matter physics. We report in this paper on reproduced observation of the magnetic flux trapping, one of the key manifestations of superconductivity, in fine air-annealed graphite powders. Trapped flux persists up to a temperature of 390 K and even above. TEM study of the un-annealed, vacuum-annealed and air-annealed samples disclosed a unique morpho-structural feature of the latter – a development of dense arrays of wrinkled regions at the surface of graphite nanoflakes. Therefore, we relate the magnetic flux trapping and possible local (granular) high-temperature superconductivity to these strained (wrinkled) regions. Furthermore, a procedure of air-annealing at moderate (670 K) temperature is shown to be an effective approach for preparation of graphite nanopowders rich with wrinkled regions at the basal flake surfaces.

**4. Experimental Section/Methods**

Thin graphite nanoflakes were obtained from bulk highly oriented pyrolytic graphite wafer, Grade ZYA, mosaic spread 0.5 ± 0.1° (Advanced Technical Centre, Moscow, Russia) via an extended grinding process. Importantly, dry grinding was carried out in an inert atmosphere of ultrapure argon or nitrogen. Grinding with lubrication with organic solvents (isopropyl alcohol or hexane) didn't lead to a reported in the paper phenomenon. To prevent oxidation and contamination, the bulk HOPG was exfoliated for 30 hours using a brand-new agate mortar. Thereafter, we took some portion of the "as-prepared" graphite flakes and annealed it in the ambient atmosphere ("air-annealed") for 24 hours at a mild temperature of 670 K. Another portion of the as-prepared powder was annealed in the vacuum at a pressure below $10^{-5}$ mbar (referred in the paper as "vacuum-annealed"). Upon completion of the annealing process, the samples were cooled down to the room temperature, maintaining their respective atmospheres. Here the chosen intermediate annealing temperature (~ 670 K) lies within the critical range that on one hand enables thermally activated local structural rearrangements in thin few-layer graphite/graphene [27, 35-36] and, on the other hand, minimizes surface degradation and oxidation (gasification reaction) effects [37].

Post-synthesis XRD and X-ray photoelectron spectroscopy (XPS) confirmed the effectiveness of this approach, revealing a high-purity carbon structure with only trace amounts of adsorbed oxygen and no significant metallic or other chemical impurities (see Supplementary Material, **Figure S1**).

To investigate the morphology of the prepared samples, we utilized scanning electron microscopy (SEM) using the MERLIN system by Carl Zeiss, which was equipped with an energy-dispersive spectrometer AZTEC X-MAX from Oxford Instruments. For high-resolution imaging and further structural analysis, we employed the Hitachi HT7700 system for transmission electron microscopy (TEM), operating at an acceleration voltage of 100 kV, with a maximum achievable resolution of 1.44 Å. This system is equipped with the selected area electron diffraction (SAED) option. The SEM results indicated that the thickness of the prepared flakes ranged from 10 to 50 nm, with lateral dimensions varying between 1 and 10 μm.

Magnetization measurements were performed using vibrating sample magnetometry (VSM) option of the Physical Property Measurement System (PPMS, Quantum Design, USA). Prior to each measurement sequence, the sample was mounted to a standard sample holder, and the PPMS solenoid was meticulously demagnetized. This was achieved by cycling the magnetic field at an elevated temperature of 380 K, well above the anticipated transition temperatures of any magnetic phases in the sample. Subsequently, the field was ramped down to zero in the so-called "oscillating" mode, a procedure designed to eliminate any trapped moment in the superconducting coil itself. This rigorous protocol was essential for obtaining reliable and reproducible zero-field-cooled (ZFC) and field-cooled (FC) magnetization data. The mass of each sample for magnetization measurement was in the range of 16 – 20 mg. An accessible temperature range was 4.2 – 400 K.

**Acknowledgements**

The authors are grateful to Dr. V. Evtugyn from Interdisciplinary Centre for Analytical Microscopy of Kazan Federal University for transmission electron microscopy image measurements.

**Data Availability Statement**

Raw data is available on reasonable request to corresponding authors.

# Supporting Information

for the article

**Wrinkles and Magnetic Flux Trapping in Pyrolytic Graphite Nanoflakes: Possible Source and Manifestation of Room Temperature Superconductivity**

by

*Muhammad Saad, Sergey I. Nikitin, Dmitry A. Tayurskii, Roman V. Yusupov*

XRD and XPS analyses (**Figure S1**) were performed to verify structural integrity and chemical purity o the graphite nanoflakes before and after annealing. The dominant (002) and (004) reflections remain intact across all conditions, indicating that the basal plane stacking order is largely preserved. Complementary XPS measurements further confirm the high purity carbon structure, with spectra primarily dominated by the C1s peak. A minor oxygen contribution remains limited and does not influence on the overall carbon framework. The absence of significant impurity-related peaks (e.g., metallic or extrinsic elements) indicates that the preparation and annealing processes do not introduce detectable contamination.

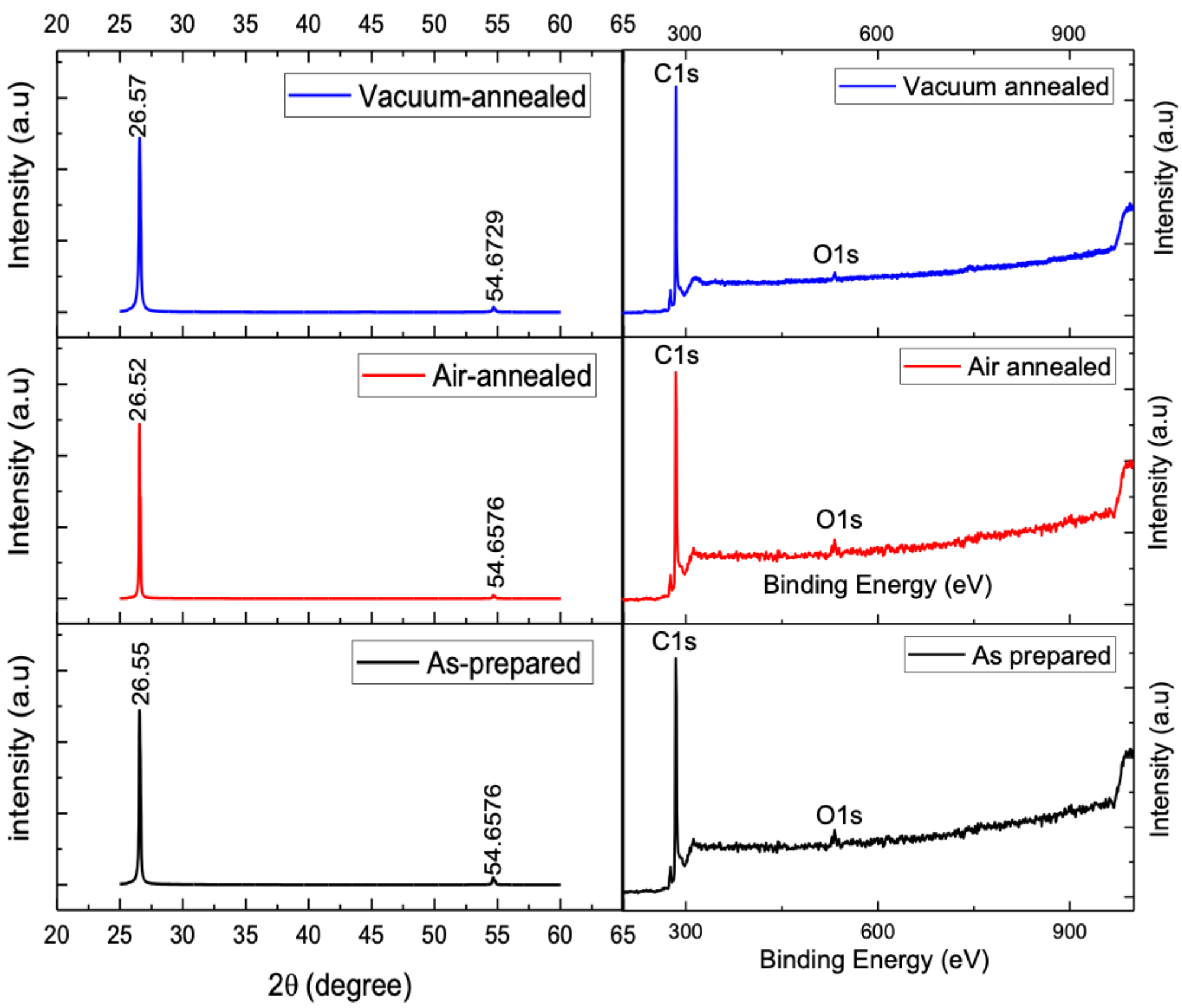

**Figure S1** – XRD diffraction patterns (left column) showing the characteristic graphitic reflections indexed to the main (002) and (004) planes. X-ray photoelectron spectroscopy (XPS) survey spectra of the three samples (right column).

To carefully identify the morphological rearrangement on the basal planes of the prepared graphite samples, we applied frequency-domain filtering to the TEM images using 2D-FFT approach. First, the extracted region from the TEM was transformed into reciprocal space, where selected spatial frequencies corresponding to the lattice periodicity was examined by masking and removing the background noise. An inverse FFT was then performed to reconstruct the real-space images with suppressed high-frequency noise. Both the raw and FFT-filtered images of the as-prepared (**Figure S2**), vacuum-annealed (**Figure S3**) and air-annealed (**Figure S4**) are displayed. The images become significantly clearer only after frequency-domain processing, allowing reliable visualization of morphological features at the basal plane of the thin graphite regions. The reconstructed image of the as-prepared sample shows relatively uniform fringe contrast, with no well-defined long-range corrugation. Similarly, under vacuum annealing the unfiltered images exhibit diffuse, low-contrast textures without well-defined corrugation. After filtering and FFT reconstruction, the contrast becomes more uniform rather than developing ordered or modulations. No periodic wrinkle-type features are resolved, indicating the absence of coherent rearranged structures. In contrast, the air-annealed sample exhibits laterally extended contrast modulations after FFT-based noise filtering. The analysis is crucial because it distinguishes true structural corrugations from imaging noise and contrast artifacts, providing robust evidence that mild annealing in the air induced wrinkle formation at the basal plane of graphite nanoflakes. This confirms that the observed features originate from real microstructural rearrangements rather than imaging artifacts.

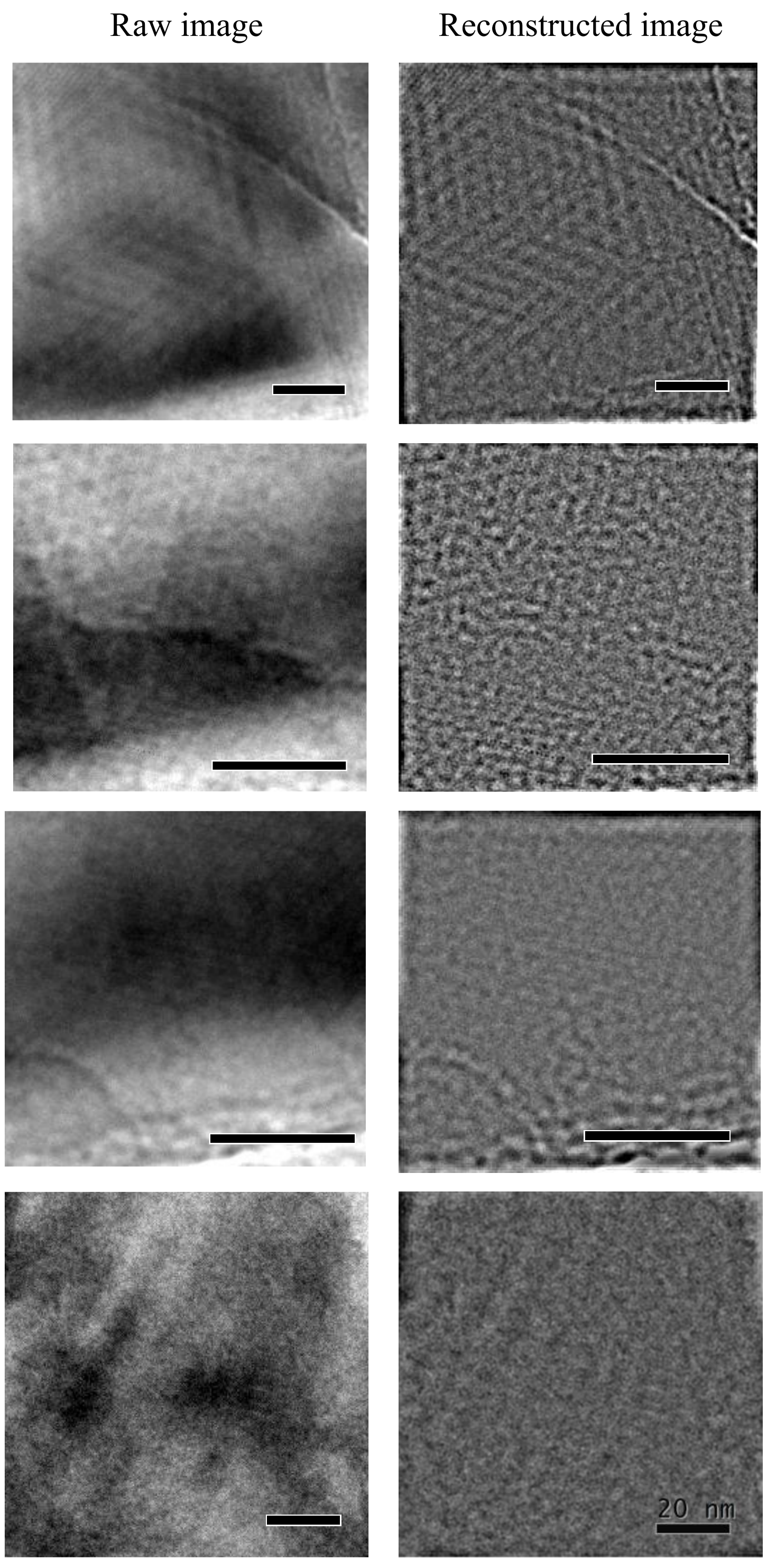

**Figure S2**: Raw (left column) and FFT-filtered (right column) basal plane TEM images of the as-prepared graphite sample.

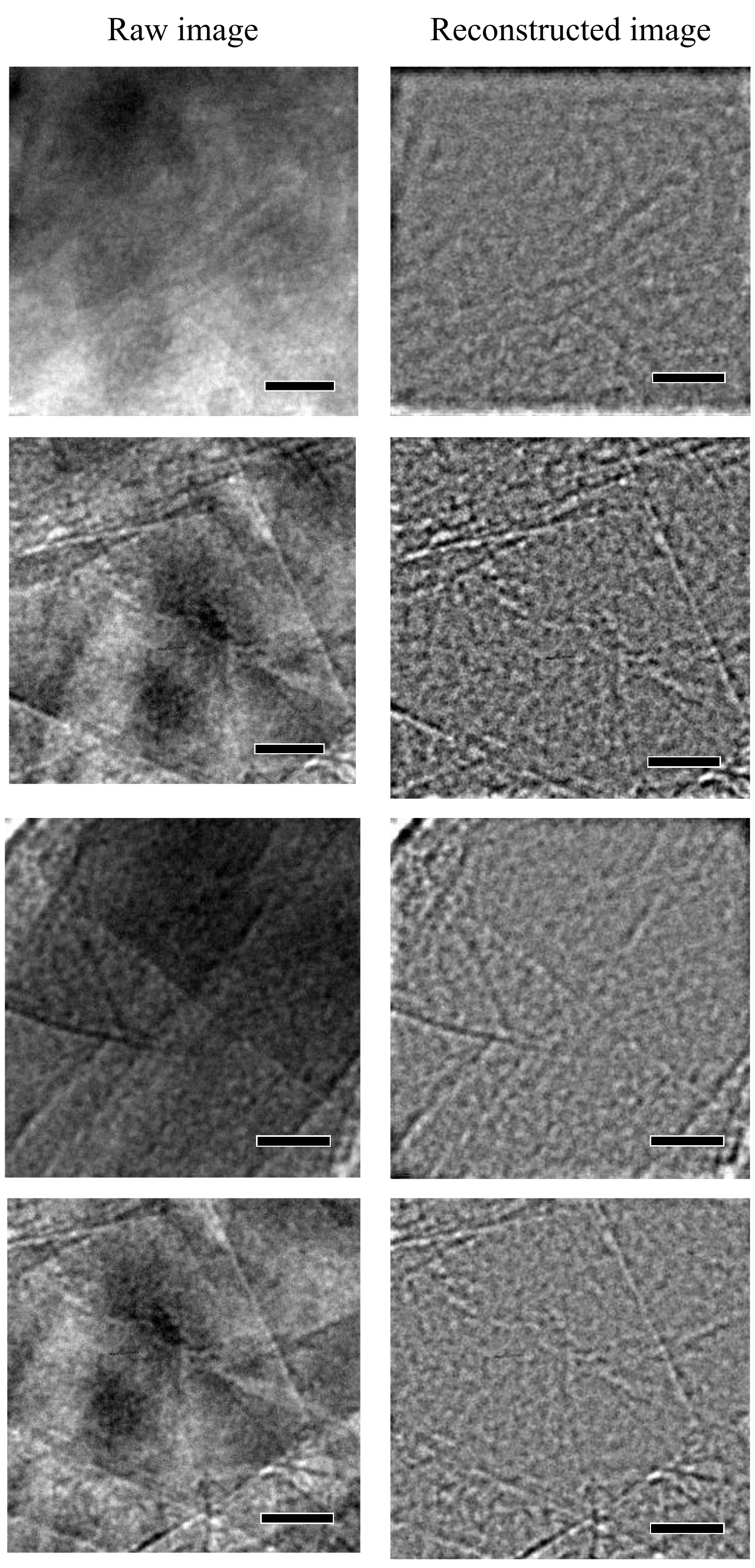

**Figure S3**: Raw (left column) and FFT-filtered (right column) basal plane TEM images of the vacuum-annealed graphite sample.

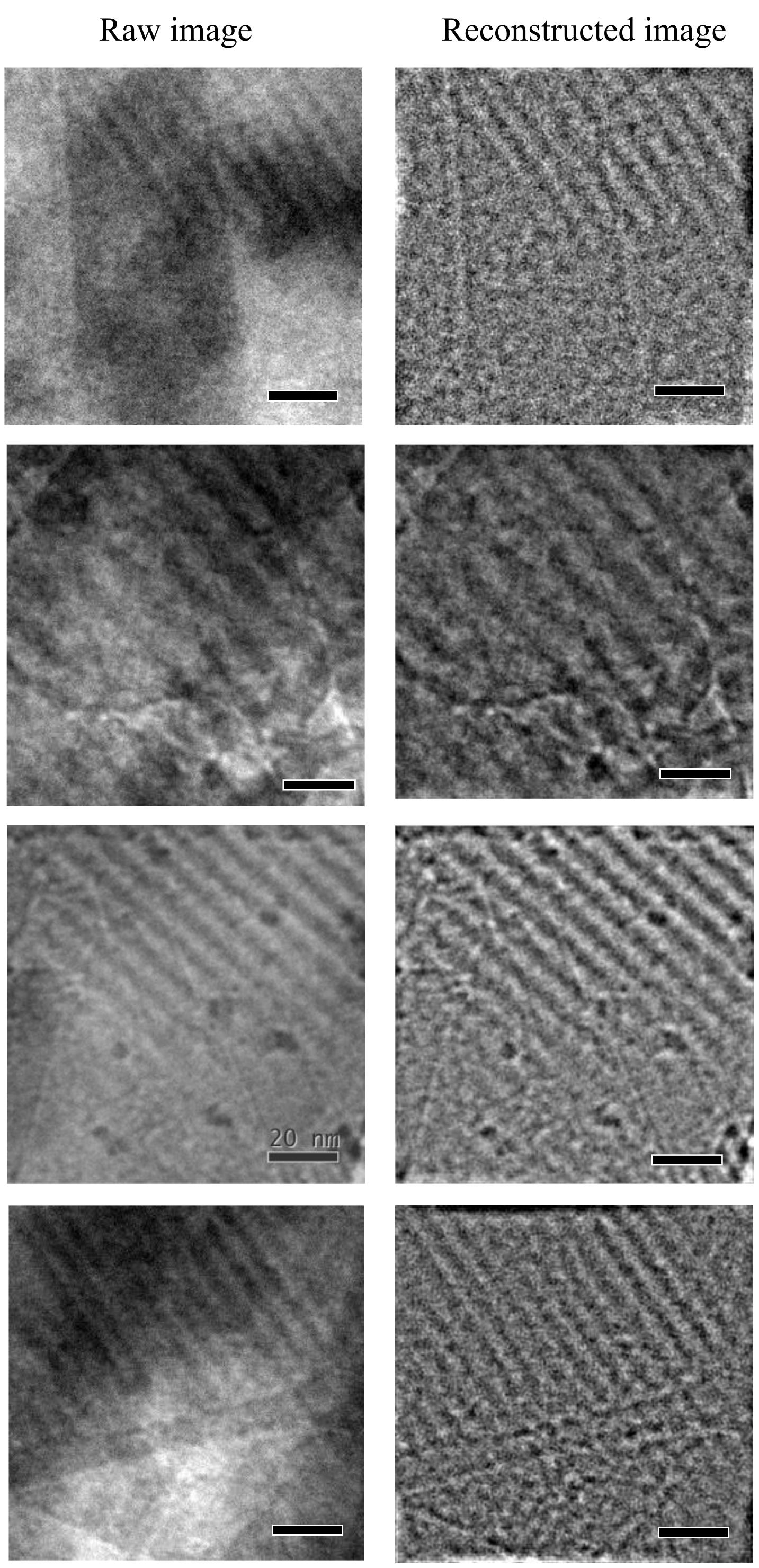

**Figure S4**: Raw (left column) and FFT-filtered (right column) basal plane TEM images of the air-annealed graphite sample.

**Field cooled (FC) M(H) data analysis of the sample annealed in the air:**

The foremost morphological analysis shows that air annealing generates dense wrinkle networks of different periodicity distributed at the graphite basal planes. And namely the air-annealed sample of a series of three graphite powders initially of the same origin manifests the magnetic anomaly that we consider as a possible magnetic flux trapping – one of the core indications of the superconductivity.

In **Figure S5**, the M(H) remagnetization curves after the field-cooling are subjected to a simple graphic analysis disclosing its assignment to one of the two known displacement origins: a) magnetic flux trapping that should manifest itself in a vertical (along a magnetic moment axis) shift, and b) exchange bias [ ] taking place at a boundary of a ferro- and antiferromagnets that results in a horizontal (along the field axis) data shifts. Therefore, we have mutually translated the original measured curves. To see whether the FC displacement arises from an exchange bias or trapped flux, each FC loop was translated independently in the field and in magnetization direction, respectively. Horizontal translation fails to recover symmetry, demonstrating that the FC curves cannot be aligned with ZFC curve by a field shift. In contrast, vertical translation produces complete overlap of the FC with ZFC curve, restoring full symmetry. The symmetric recovery of the FC curves after vertical translation, and their collapse onto the ZFC curve demonstrate that the system rather traps flux than exhibits a field-shifted (exchange-biased) response.

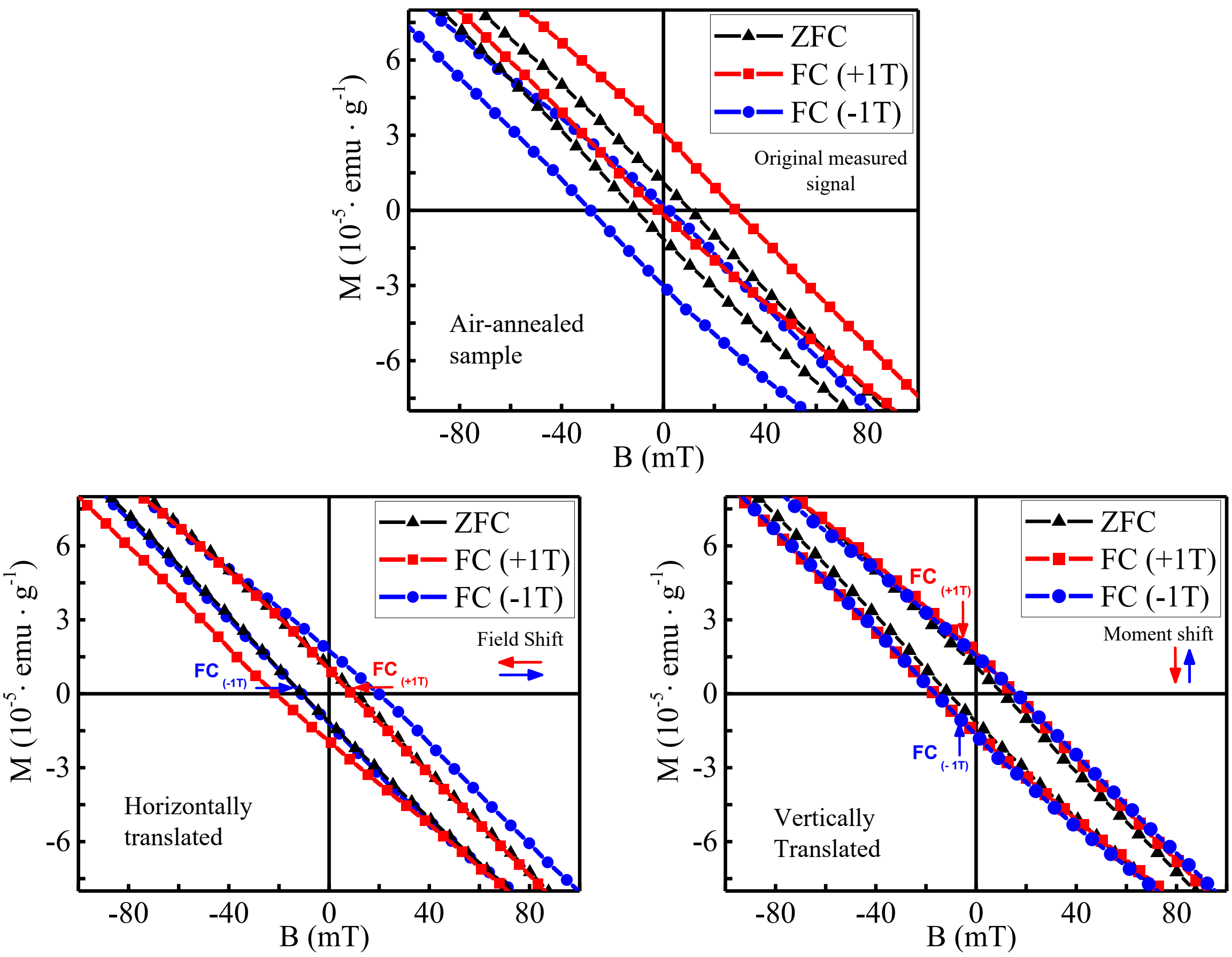

**Figure S5**: Analysis of ZFC and FC hysteresis loops of the air-annealed graphite powder. Each panel shows the original measured magnetization curve at 10 K under zero-field-cooled (ZFC) protocols: i) 0 → +1T → -1T → +1T (black trace), providing an independent virgin loop. Field-cooled FC curves were recorded after cooling from 350 K to 10 K under +1T (red trace) or -1T (green trace). The right bottom panel shows the vertically translated FC hysteresis loops confirming their perfect overlap.